%% file: VTC_on_MaMi_Mobility.tex
\documentclass[conference]{IEEEtran}
\usepackage[T1]{fontenc}

  \usepackage[pdftex]{graphicx}
  \graphicspath{{./png/}{./pdf/}{./jpeg/}}
  \DeclareGraphicsExtensions{.pdf,.jpeg,.png}
\usepackage{amsmath}
\usepackage[cmintegrals]{newtxmath}
\ifCLASSOPTIONcompsoc
  \usepackage[caption=false,font=normalsize,labelfont=sf,textfont=sf]{subfig}
\else
  \usepackage[caption=false,font=footnotesize]{subfig}
\fi

\usepackage{stfloats}
\usepackage{todonotes}
\usepackage[detect-weight=true, detect-family=true, per-mode=symbol]{siunitx}
\usepackage{booktabs}
\renewcommand{\vec}[1]{\ensuremath{\boldsymbol{#1}}}

\usepackage[acronym, nomain,nogroupskip,nonumberlist,nopostdot, toc]{glossaries}
\loadglsentries[\acronymtype]{acronyms}
\begin{document}
\bstctlcite{IEEEexample:BSTcontrol}
%
\title{Temporal Analysis of Measured LOS Massive MIMO Channels with Mobility}

\author{\IEEEauthorblockN{Paul Harris\IEEEauthorrefmark{1},
Steffen Malkowsky\IEEEauthorrefmark{2},
Joao Vieira\IEEEauthorrefmark{2}, 
Fredrik Tufvesson\IEEEauthorrefmark{2},
Wael Boukley Hasan\IEEEauthorrefmark{1},
Liang Liu\IEEEauthorrefmark{2}\\
Mark Beach\IEEEauthorrefmark{1},
Simon Armour\IEEEauthorrefmark{1} and
Ove Edfors\IEEEauthorrefmark{2}}

\IEEEauthorblockA{\IEEEauthorrefmark{1}\gls{CSN} Group, University of Bristol, U.K. \\Email: \{paul.harris, wb14488, m.a.beach, simon.armour\}@bristol.ac.uk\
\IEEEauthorblockA{\IEEEauthorrefmark{2}Dept. of Electrical and Information Technology, Lund University, Sweden\\
Email: firstname.lastname@eit.lth.se}}}

\maketitle

\begin{abstract}
The first measured results for massive \gls{MIMO} performance in a \gls{LOS} scenario with moderate mobility are presented, with 8 users served by a 100 antenna \gls{BS} at \SI{3.7}{\giga\hertz}. When such a large number of channels dynamically change, the inherent propagation and processing delay has a critical relationship with the rate of change, as the use of outdated channel information can result in severe detection and precoding inaccuracies. For the \gls{DL} in particular, a \gls{TDD} configuration synonymous with massive \gls{MIMO} deployments could mean only the \gls{UL} is usable in extreme cases. Therefore, it is of great interest to investigate the impact of mobility on massive \gls{MIMO} performance and consider ways to combat the potential limitations. In a mobile scenario with moving cars and pedestrians, the correlation of the \gls{MIMO} channel vector over time is inspected for vehicles moving up to \SI{29}{\kilo\meter\per\hour}. For a 100 antenna system, it is found that the \gls{CSI} update rate requirement may increase by 7 times when compared to an 8 antenna system, whilst the power control update rate could be decreased by at least 5 times relative to a single antenna system. 
\end{abstract}

%
\IEEEpeerreviewmaketitle

\section{Introduction}
\glsresetall
%
%
%
%
\IEEEPARstart{M}{assive} \gls{MIMO} has established itself as a key 5G technology that could drastically enhance the capacity of sub-\SI{6}{\giga\hertz} communications in future wireless networks. By taking the \gls{MU} \gls{MIMO} concept and introducing additional degrees of freedom in the spatial domain, the multiplexing performance and reliability of such systems can be greatly enhanced, allowing many tens of users to be served more effectively in the same time-frequency resource \cite{Marzetta2010}. 
In addition to theoretical work and results such as those documented in \cite{Marzetta2010}, \cite{6736761}, \cite{6891254} and \cite{Hoydis2013}, various institutions from around the world have been developing large-scale test systems in order to validate theory and test algorithms with real data. Some known examples include the Rice University Argos system \cite{Shepard2012} \cite{Shepard2013}, the Ngara demonstrator in Australia \cite{Suzuki2012a}, the \gls{FD-MIMO} work by Samsung \cite{6810440} \cite{6525612}, field trials by ZTE \cite{Ux2015} and Project Aries by Facebook \cite{Facebook}. The work presented here is underpinned by 100-antenna and 128-antenna real-time testbeds developed by Lund University and the University of Bristol in collaboration with National Instruments \cite{Vieira2014a} \cite{Malkowsky2016b}. In 2016, two indoor trials were conducted within an atrium at the University of Bristol, and it was shown that spectral efficiencies of 79.4 bits/s/Hz and subsequently 145.6 bits/s/Hz could be achieved whilst serving up to 22 user clients \cite{PaulGlobecom} \cite{PaulSIPS}. These spectral efficiency results are currently world records and indicate the potential massive \gls{MIMO} has as a technology.


However, wireless technology is usually applied in scenarios that will involve some form of mobility, and it is therefore of great interest to investigate the evolution of massive \gls{MIMO} channels under more dynamic conditions. The aforementioned measurement trials have not yet considered the progression of a composite massive \gls{MIMO} channel with mobility, but measured static terminals. In \cite{7510708}, the authors discuss some of the potential issues with channel ageing in a macro-cell massive \gls{MIMO} deployment and propose a novel \gls{UDN} approach for comparison. From simulation results, they illustrate that massive \gls{MIMO} performance could significantly worsen with mobile speeds of just \SI{10}{\kilo\meter\per\hour}, and that the sensitivity of \gls{ZF} to \gls{CSI} errors could make \gls{MF} the more viable option. In \cite{7519076} and \cite{7473866}, the theoretical impact of channel ageing on the \gls{UL} and \gls{DL} performance of massive \gls{MIMO} is evaluated. Interestingly, the analysis shows that a large number of antennas is to be preferred for maximum performance, even under time-varying conditions, and that Doppler effects dominate over phase noise.

In this paper, 8 \glspl{UE} are served by a 100 antenna \gls{BS} in real-time and the correlation over time for both the power and \gls{MIMO} channel is measured for speeds of up to \SI{29}{\kilo\meter\per\hour}. To the best of the authors' knowledge, these are the first measured results for massive \gls{MIMO} under moderate mobility in \gls{LOS}.
\section{System Description}
\label{System}
\begin{figure}[!t]
	\centering
	\includegraphics[width=0.75\columnwidth]{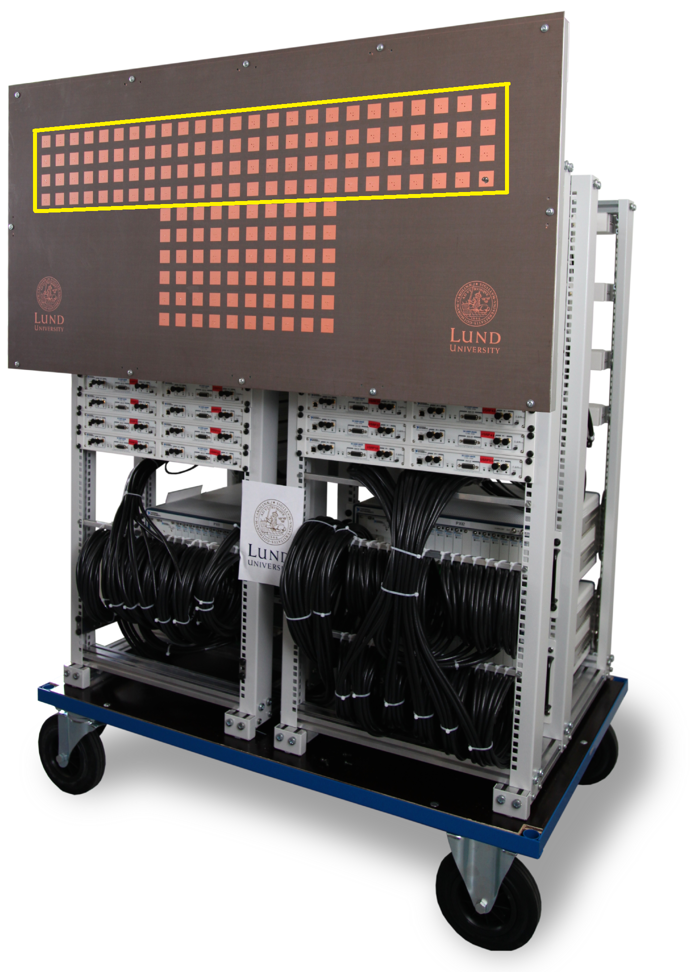}
	\caption{Lund Massive MIMO Basestation with 4x25 array configuration highlighted}
	\label{fig_BS}
\end{figure}

The Lund University massive \gls{MIMO} \gls{BS} pictured in ~\figurename~\ref{fig_BS} consists of 50 \gls{NI} \glspl{USRP}, which are dual-channel \glspl{SDR} with reconfigurable \glspl{FPGA} connected to the \gls{RF} front ends \cite{USRP}. Collectively, these provide 100 \gls{RF} chains, with a further 6 \glspl{USRP} acting as 12 single-antenna \glspl{UE}. It runs with a \gls{TDD} LTE-like \gls{PHY} and the key system parameters can be seen in Table \ref{tab:system_param}. Using the same \gls{PXIe} platform that powers the 128-antenna system, all the \glspl{RRH} and \gls{MIMO} \gls{FPGA} co-processors are linked together by a dense network of gen 3 \gls{PCIe} fabric,
and all software and \gls{FPGA} behaviour is programmed via LabVIEW. A description of the Lund University system along with a general discussion of massive \gls{MIMO} implementation issues can be found in \cite{Malkowsky2016b}. 
The reciprocity calibration approach designed at Lund University and applied in these experiments is detailed in \cite{VieiraREMLT16}. Further detail about the current system architecture and the implementation of a wide data-path \gls{MMSE} encoder/decoder can be found in \cite{PaulGlobecom}, \cite{PaulSIPS} and \cite{7780107}.
\begin{table}
	\renewcommand{\arraystretch}{1.3}
	\caption{System Parameters}
	\centering
	\noindent\begin{tabular}{ll}
		\toprule
		\textbf{Parameter} & \textbf{Value}  \\
		\midrule
		\# of BS Antennas & 100  \\
		\# of UEs & 12 \\
		Carrier Frequency & 1.2-\SI{6}{\giga\hertz} (\SI{3.7}{\giga\hertz} used) \\
		Bandwidth		  & \SI{20}{\mega\hertz}  \\
		Sampling Frequency  & \SI{30.72}{\mega S\per\second} \\
		Subcarrier Spacing  & \SI{15}{\kilo\hertz} \\
		\# of Subcarriers  &  2048 \\
		\# of Occupied Subcarriers  & 1200 \\
		Frame duration  & \SI{10}{\milli\second} \\
		Subframe duration & \SI{1}{\milli\second} \\
		Slot duration & \SI{0.5}{\milli\second} \\
		TDD periodicity &   1 slot  \\	
		\bottomrule
	\end{tabular}
	\label{tab:system_param}
\end{table}
\subsection{Antenna Array}
The Lund antenna array in~\figurename~\ref{fig_BS} consists of half-wavelength spaced 3.7 GHz patch-antennas each with horizontal and vertical polarization options. For the results reported here, the 4x25 azimuth dominated portion highlighted in the image was used, with alternating horizontal and vertical polarizations across the array.

\subsection{Channel Acquisition}
The system is defined as having $M$ antennas, $K$ users and $N$ frequency domain resource blocks of size $R_{b}$. For this paper, we specifically define resource block to refer to a grouping of 12 \gls{OFDM} subcarriers, with each 1200 subcarrier \gls{OFDM} symbol consisting of 100 resource blocks.
The estimate of the \gls{UL} for each 12 subcarrier resource block $r$ is found as
\begin{equation}
\vec{H}_{r} = \vec{Y}_{r}\vec{P}^{*}_{r}
\end{equation}
where $\vec{H}_{r}$ is the $M \times K$ channel matrix, $\vec{Y}_{r}$ is the $M \times R_{b}$ receive matrix and $\vec{P}^{*}_{r}$ is the diagonal $R_{b} \times K$ conjugate uplink pilot matrix. Each \gls{UE} sends an \gls{UL} pilot for each resource block on a subcarrier orthogonal to all other users and the \gls{BS} performs least-square channel estimation. Since all pilots have a power of 1, $\vec{Y}_{r}\vec{P}^{*}_{r}$ can be used rather than $\vec{Y}_{r} / \vec{P}^{*}_{r}$.
All \gls{MIMO} processing in the system is distributed across 4 Kintex 7 \gls{FPGA} co-processors, each processing 300 of the 1200 subcarriers. In order to study the channel dynamics under increased levels of mobility, a high time resolution is desired during the capture process, which equates to a high rate of data to write to disk. To address this, a streaming process was implemented that uses the on-board \gls{DRAM} to buffer the raw \gls{UL} subcarriers.
$\vec{Y}_{r}$ is recorded in real-time to the \gls{DRAM} at the measurement rate, $T_\text{meas}=\SI{5}{\milli\second}$, and this data is then siphoned off to disk at a slower rate. 
Using this process with \SI{2}{\giga B} of \gls{DRAM} per \gls{MIMO} processor, we were able to capture the full composite channel for all resource blocks a \SI{5}{\milli\second} resolution for \SI{65}{\second}.
\\

The channel was sampled at least once every half-wavelength distance in space to give an accurate representation of the environment. The maximum permissible speed of mobility, $v_\text{max}$, is thus given by
\begin{equation}
v_\text{max} = \frac{\lambda}{2T_\text{meas}}.
\label{speed}
\end{equation}
This results in a maximum speed of \SI{8.1}{\meter\per\second} or approximately \SI{29}{\kilo\meter\per\hour} for temporal analysis of the channel data in this case. 

\subsection{Post-Processing}
To evaluate the change in the multi-antenna channels under mobility, an analog of the \gls{TCF} \cite{molisch2010wireless} was calculated. By introducing a time dependence on the measured channels, i.e., the channel vector corresponding to user $i$ at resource block $r$ and at time $t$ is denoted by $\vec{h}_{i,r}[t]$, the \gls{TCF} was defined as
\begin{equation}
\text{TCF}_{i}(\tau) = \frac{\mathbf{E}\{ | \vec{h}_{i,r}[t-\tau]^H \; \vec{h}_{i,r}[t] | \} }{ \mathbf{E}\{ \vec{h}_{i,r}[t]^H  \; \vec{h}_{i,r}[t] \}},
\end{equation}
where $\mathbf{E}\{ \}$ denotes the expectation operator. At a given time lag $\tau$, the expectation is computed according to its definition, but also by averaging over all resource blocks for better statistics.


\subsection{User Equipment}
\begin{figure}[!t]
	\centering
	\includegraphics[width=\columnwidth]{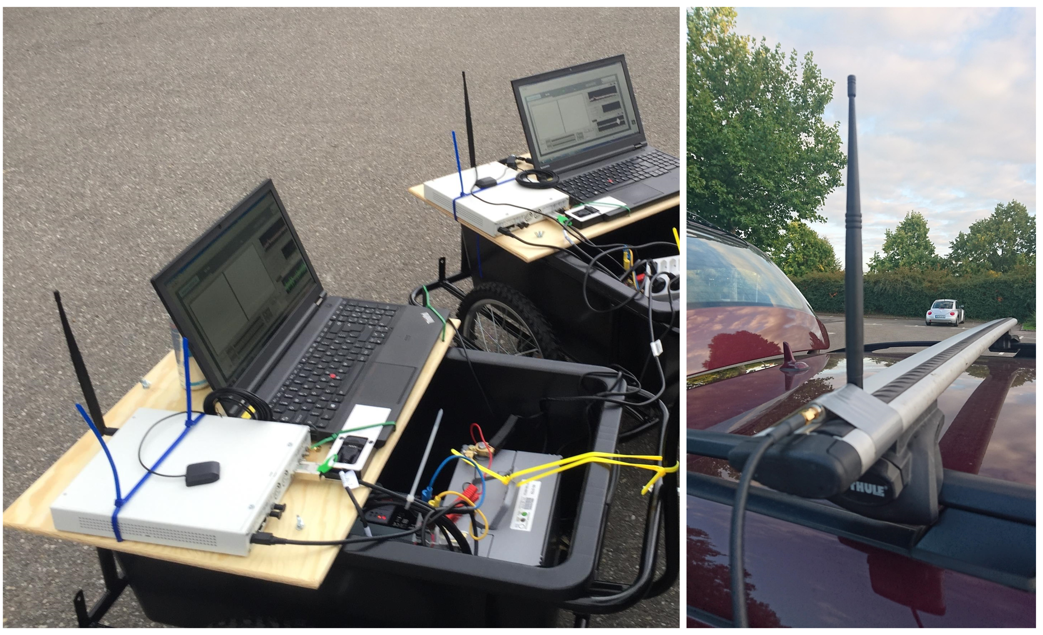}
	\caption{User Equipments. Left: pedestrian carts. Right: car mounting.}
	\label{fig_UEs}
\end{figure}

As mentioned at the beginning of Sec.~\ref{System}, each \gls{UE} is a two-channel \gls{USRP}, four of which were used in these measurements to provide a total of 8 spatial streams.
The \glspl{USRP} were mounted in carts to emulate pedestrian behaviour and in cars for higher levels of mobility, as shown in ~\figurename~\ref{fig_UEs}.
Sleeve dipole antennas where used in each case. On the carts, the dipole antennas were attached directly to both \gls{RF} chains of the \gls{USRP} with vertical polarisation. With a fixed spacing of 2.6$\lambda$, these can either be considered as two devices in very close proximity or as a single, dual-antenna device. In~\figurename~\ref{fig_UEs}, the \glspl{USRP} shown in carts each have only one antenna connected.
For the cars, the antennas were roof mounted with vertical polarization on either side of the car, giving a spacing of approximately \SI{1.7}{\meter}. Each \gls{UE}'s
\gls{RF} chain operates with a different set of frequency-orthogonal pilots and synchronises \gls{OTA} with the \gls{BS} using the \gls{PSS} broadcast at the start of each \SI{10}{\milli\second} frame. The \gls{PSS} was transmitted using a static beam pattern and the \glspl{LO} of the \glspl{UE} were \gls{GPS} disciplined to lower carrier frequency offset, thereby improving stability during the measurements.

\section{Measurement Scenario}
\label{Scen}
\begin{figure}[!t]
	\centering
	\includegraphics[width=\columnwidth]{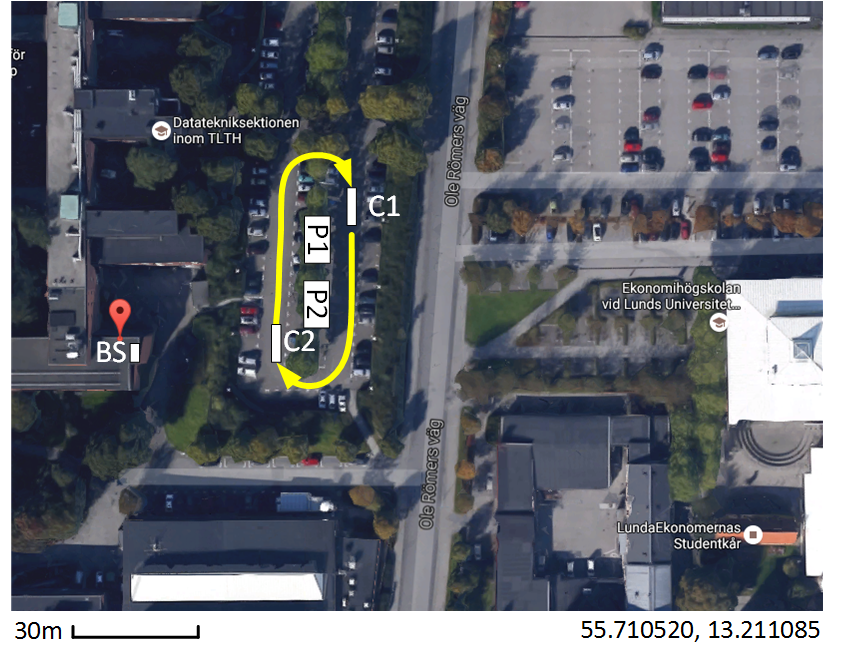}
	\caption{Overview of the measurement scenario at the campus of the Faculty of Engineering (LTH), Lund University, Sweden. Arrows indicate the direction of movement for the cars C1 and C2. Pedestrians P1 and P2 moved within the zones indicated by their white boxes.}
	\label{fig_mob_above}
\end{figure}

A mixture of both pedestrian and vehicular \glspl{UE} were used so that it would be possible to observe how the massive \gls{MIMO} channel behaves over time in a more dynamic situation. Each user transmitted with the same fixed power level and $\vec{Y}_{r}\forall \ r$ was captured for a \SI{30}{\second} period. Two pedestrian carts, indicated by P1 and P2, moved pseudo-randomly at walking pace to and from one another for the measurement duration, whilst two cars, shown as C1 and C2, followed the circular route shown. For the temporal results concerning the cars, it was ensured that the captures analysed were from a period of the scenario where the cars did not exceed our maximum $\lambda / 2$ measurement speed of \SI{29}{\kilo\meter\per\hour}. Over the course of the entire capture, the cars completed approximately two laps and arrived back at the starting position indicated in ~\figurename~\ref{fig_mob_above}. With the cars moving in this pattern, the devices are, on average, more distributed in the azimuth, but when C1 and C2 pass in parallel to the pedestrian carts they become more clustered in a perpendicular line to the \gls{BS}.

\section{Results}
\label{Res}
\begin{figure}[!t]
	\centering
	\subfloat[]{\includegraphics[width=\columnwidth]{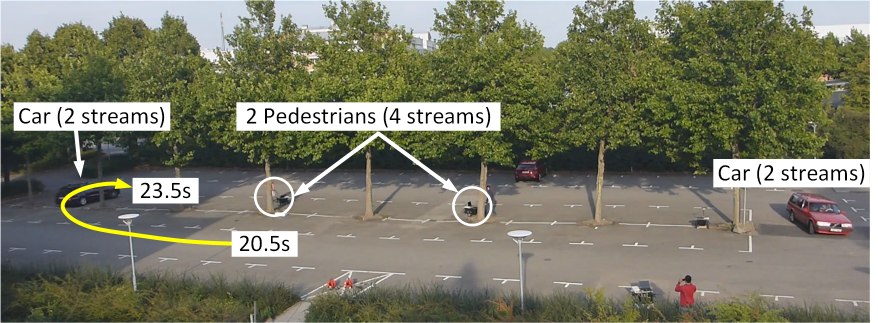}}\\
	\subfloat[]{\includegraphics[width=\columnwidth]{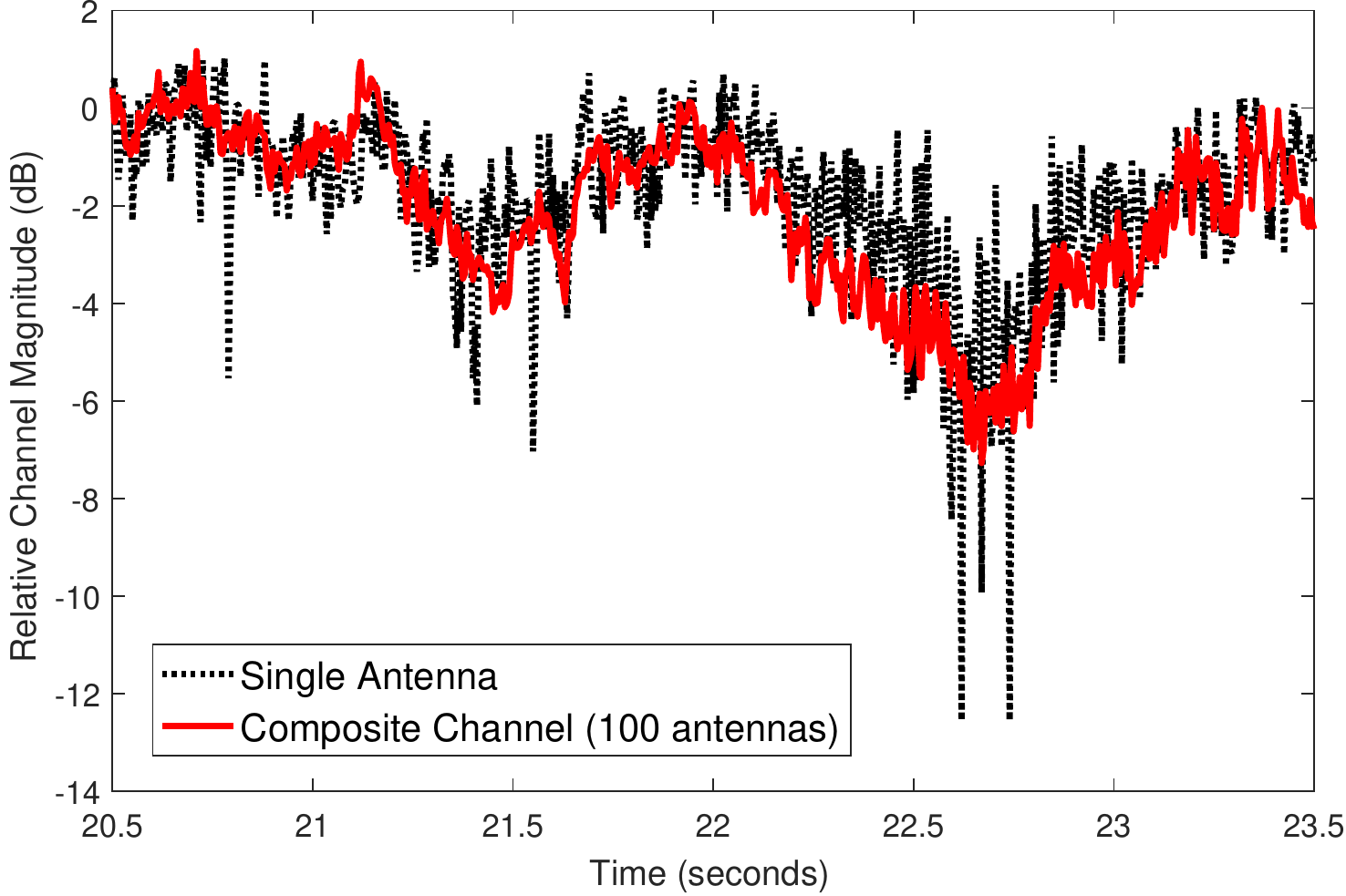}}
	\caption{Resilience to fading. a) View from \gls{BS} with 3 second car-based user path indicated. b) Relative channel magnitude for both a single antenna and the composite \gls{MIMO} channel to the depicted user over the 3 second period.}
	\label{fig_fading}
\end{figure}
The temporal results were based on two different time periods within the 30 second mobility scenario. Channel hardening and power correlation results are presented first for one car user using a 3 second period of the full 30 second capture, specifically between \SI{20.5}{\second} and \SI{23.5}{\second}. Time correlation results are then shown for a 4 second period between \SI{10}{\second} and \SI{14}{\second} where both cars travel in parallel to the \gls{BS}. For both of these time periods, the vehicle speed remains below \SI{29}{\kilo\meter\per\hour}.
\\
\subsection{Power Correlation}
Fast-fading is shown to disappear theoretically when letting the number of \gls{BS} antennas go to infinity, as discussed in \cite{Marzetta2010} and \cite{6736761}.
Whilst the measured scenarios will have been more of a Rician than Rayleigh nature due to the \glspl{UE} being predominantly in \gls{LOS}, it was still possible to inspect the less severe fading dips of a single channel for a single \gls{UE} and evaluate them against the composite channel formed by the $100\times8$ massive \gls{MIMO} system. 
In \figurename~\ref{fig_fading}a, a 3 second portion of the captured mobile scenario is shown as viewed from the \gls{BS}. An arrow indicates the movement of one of the cars during this three second period. For one \gls{UE} in this car over the acquisition period shown, the channel magnitude of a single, vertically polarised \gls{BS} antenna was extracted, along with the respective diagonal element of the user side Gram matrix $\vec{H}_{r}^\mathsf{H}\vec{H}_{r}$ for one resource block $r$. Their magnitudes are plotted against each other in \figurename~\ref{fig_fading}b after normalization. It can be seen that the composite channel tends to follow the average of the single antenna case, smoothing out the faster fading extremities, and larger variations occur over the course of seconds rather than milliseconds.
\\
\begin{figure}[!t]
	\centering
	\includegraphics[width=\columnwidth]{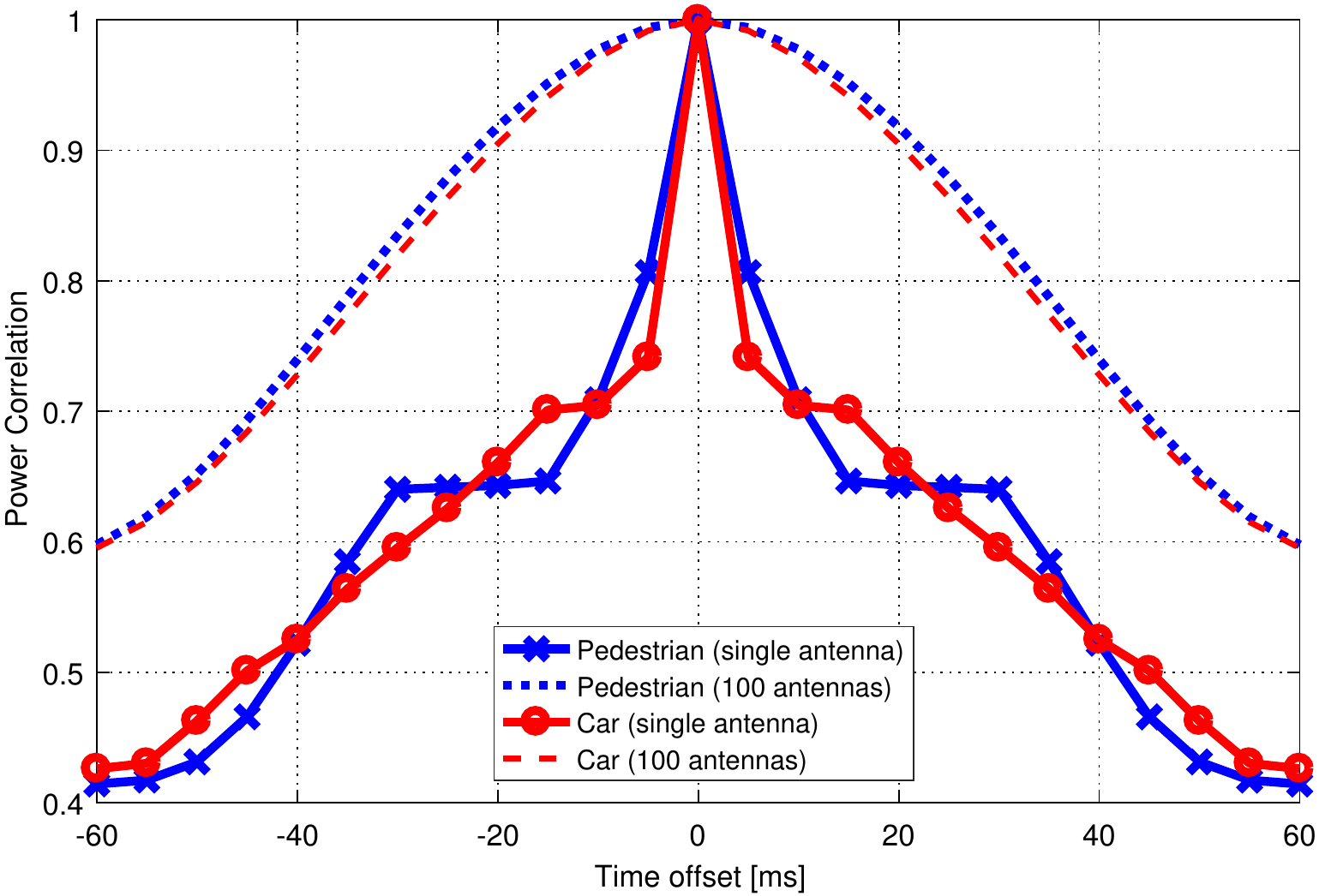}
	\caption{Correlation of the power for signal on one antenna versus 100 antennas for a pedestrian and car UE.}
	\label{fig_power_cor}
\end{figure}
\figurename~\ref{fig_power_cor} shows the correlation of the signal power over time offset for a pedestrian UE and a car UE when using either one antenna or 100 antennas.
Due to the channel hardening and the constructive combining of 100 signals, the power levels are much more stable as compared to the single antenna case. For this particular case, with 100 antennas, power control can be done at least five times slower than with a single antenna.
These two figures show, that performance of this nature not only demonstrates an improvement in robustness and latency due to the mitigation of fast-fade error bursts, but also that it is possible to greatly relax the update rate of power control when combining signals from many antennas.
\\
\subsection{Channel Correlation}
\begin{figure}[!t]
	\centering
	\includegraphics[width=\columnwidth]{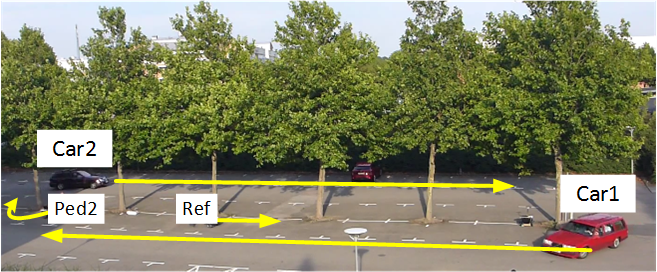}
	\caption{4 second subset of mobility scenario (\SI{10}{\second} to \SI{14}{\second} of full 30 second capture) used for temporal analysis. Arrows indicate the movement of each \gls{UE} over the \SI{4}{\second} duration. Car 2 does not exceed the maximum allowed speed of \SI{29}{\kilo\meter\per\hour}.}
	\label{fig_temp_scen}
\end{figure}
\begin{figure}[!t]
	\centering
	\includegraphics[width=\columnwidth]{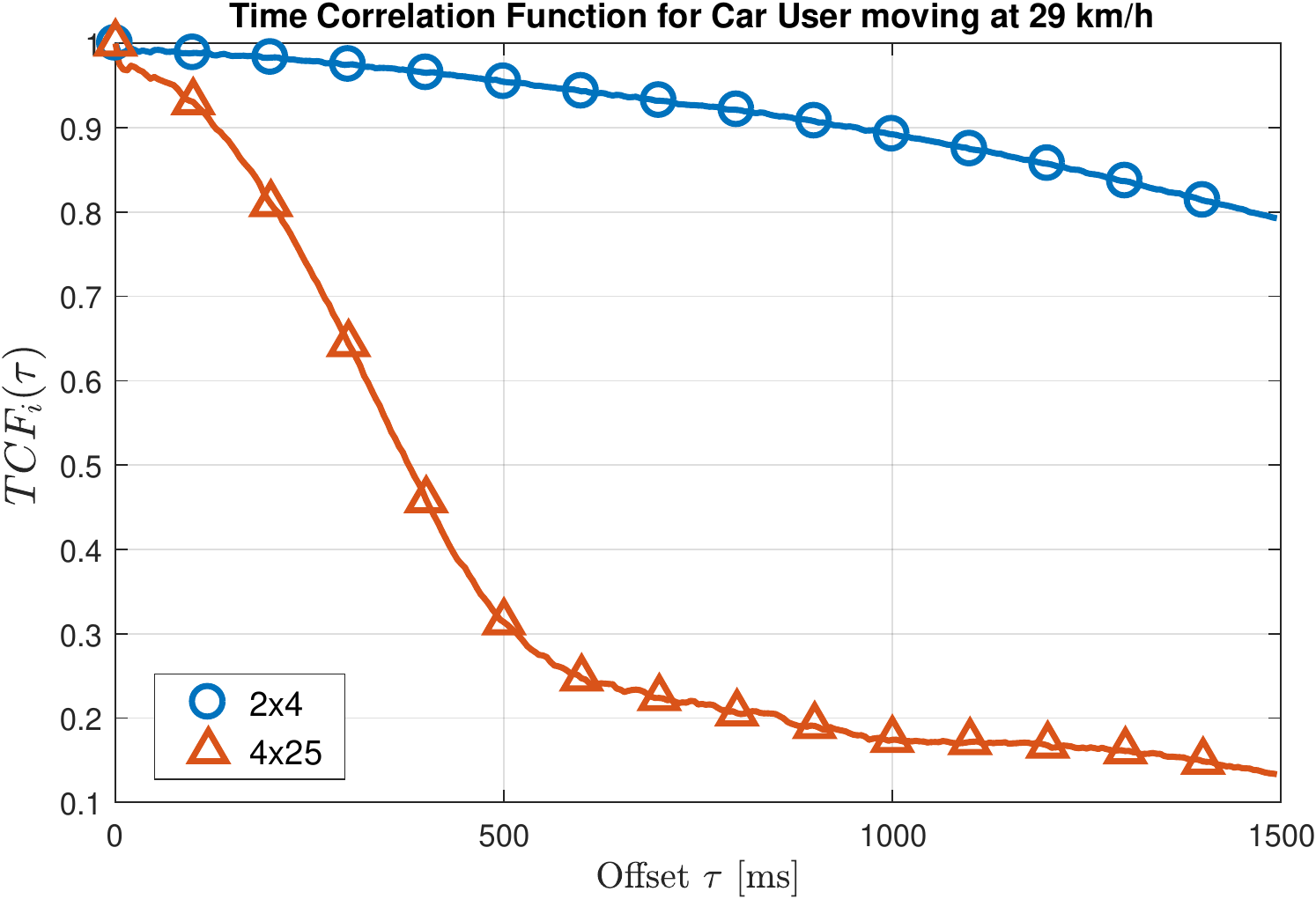}
	\caption{Correlation of the composite channel over time for all resource blocks of car 2 at a speed of \SI{29}{\kilo\meter\per\hour}. 100 antenna and 8 antenna cases are shown in 4x25 and 2x4 configurations respectively. Both curves are normalized to themselves, so array gain is not visible here.}
	\label{fig_TCF}
\end{figure}

As a \gls{UE} moves, it is of interest to view the correlation of the \gls{MIMO} channel vectors over time in order to ascertain how quickly the channel becomes significantly different. This will play a part in determining the required channel estimation periodicity for a given level of performance.
In \figurename~\ref{fig_temp_scen}, a \SI{4}{\second} second period of the 30 second mobility scenario is shown, with the arrows indicating the movement of each \gls{UE} during that period. Using one \gls{UE} from Car 2, the absolute values of the time correlation function for all resource blocks over the 4 second period are shown in \figurename~\ref{fig_TCF} for the first \SI{1.5}{\second} of movement.
Within the first \SI{500}{\milli\second} at this speed, the level of correlation has dropped significantly in the 100 antenna case to 0.3, whilst the 8 antenna case remains above 0.8 for the entire \SI{1.5}{\second} duration with a far shallower decay. For the 8 antenna and 100 antenna cases to become decorrelated by 20 percent, it takes \SI{1455}{\milli\second} and \SI{205}{\milli\second} respectively; a factor difference of approximately 7. This highlights how the precise spatial focusing of energy provided by massive \gls{MIMO} translates to a larger decorrelation in time from far smaller movements in space. The acceptable level of decorrelation will depend upon many factors such as the desired level of performance, the detection/precoding technique and the \gls{MCS}, but this result provides some insight into how rapidly a real channel vector can change in massive \gls{MIMO} under a moderate level of mobility when compared to a more conventional number of antennas.
\section{Conclusions}
\label{conc}
The temporal performance of a $100\times8$ real-time massive \gls{MIMO} system operating in a \gls{LOS} scenario with moderate mobility has been presented. To the best of the authors' knowledge, these are the first results of their kind that begin to indicate the performance of massive \gls{MIMO} as the composite channel changes over the course of a more dynamic scenario. When considering the correlation of mobile channel vectors over time, it was shown that the 100 antenna case decorrelated by 20\% 7 times faster than the 8 antenna case, providing an indication of the spatial focusing effect in a \gls{MIMO} system. In addition, the power correlation results indicate that power control algorithms may be able to operate 5 times slower than a single antenna case due to the channel hardening effect.


%

\appendices


\section*{Acknowledgment}
The authors wish to acknowledge and thank all academic staff and post graduate students involved who contributed to the measurement trial operations. They also acknowledge the financial support of the \gls{EPSRC} \gls{CDT} in Communications (EP/I028153/1), the European Union Seventh Framework Programme (FP7/2007-2013) under grant agreement no. 619086 (MAMMOET), NEC, NI, the Swedish Foundation for Strategic Research and the Strategic Research Area ELLIIT.

\ifCLASSOPTIONcaptionsoff
  \newpage
\fi



\bibliographystyle{MyIEEEtran}
\bibliography{library}
%
%


%
\begin{IEEEbiography}
	[{\includegraphics[width=1in,height=1.25in,clip,keepaspectratio]{pharris}}]{Paul Harris}
graduated from the University of Portsmouth with a 1st Class Honours degree in Electronic Engineering in 2013 and joined the Communication Systems \& Networks Group at the University of Bristol in the same year to commence a PhD. His research interests include massive \gls{MIMO} system design, performance evaluation in real-world scenarios, and the optimisation of techniques such as user grouping or power control using empirical data. Working in collaboration with Lund University and National Instruments, he implemented a 128-antenna massive MIMO test system and led two research teams to set spectral efficiency world records in 2016. For this achievement, he received 5 international awards from National Instruments, Xilinx and Hewlett Packard Enterprise, and an honorary mention in the 2016 IEEE ComSoc Student Competition for "Communications Technology Changing the World".
\end{IEEEbiography}
\begin{IEEEbiography}[{\includegraphics[width=1in,height=1.25in,clip,keepaspectratio]{Bio_Pics/Steffen.jpg}}]{Steffen Malkowsky}
	received the B.Eng. degree in Electrical Engineering and Information Technology from Pforzheim University, Germany in 2011 and the M.Sc. degree in Electronic Design from Lund University in 2013. He is currently a PhD student in the Digital ASIC group at the department of Electrical and Information Technology, Lund University. His research interests include development of reconfigurable hardware and implementation of algorithms for wireless communication with emphasis on massive MIMO. For the development of a massive MIMO testbed in collaboration with University of Bristol and National Instruments and a set spectral efficiency world record, he received 5 international awards from National Instruments, Xilinx and Hewlett Packard Enterprise.
\end{IEEEbiography}
\begin{IEEEbiography}[{\includegraphics[width=1in,height=1.25in,clip,keepaspectratio]{Bio_Pics/Joao.jpg}}]{Joao Vieira}
	received the B.Sc. degree in Electronics and Telecommunications Engineering from University of Madeira in 2011, and the M.Sc. degree in Wireless Communications from Lund University, Sweden in 2013. He is currently working towards a Ph.D. degree at the department of Electrical and Information Technology in Lund University. His main research interests regard parameter estimation and implementation issues in massive MIMO systems.
\end{IEEEbiography}
\begin{IEEEbiography}[{\includegraphics[width=1in,height=1.25in,clip,keepaspectratio]{Bio_Pics/Tufvesson.jpg}}]{Fredrik Tufvesson}
	received his Ph.D. in 2000 from Lund University in Sweden. After two years at a startup company, he joined the department of Electrical and Information Technology at Lund University, where he is now professor of radio systems. His main research interests are channel modelling, measurements and characterization for wireless communication, with applications in various areas such as massive MIMO, UWB, mm wave communication, distributed antenna systems, radio based positioning and vehicular communication. Fredrik has authored around 60 journal papers and 120 conference papers, recently he got the Neal Shepherd Memorial Award for the best propagation paper in IEEE Transactions on Vehicular Technology.
\end{IEEEbiography}
\begin{IEEEbiography}
	[{\includegraphics[width=1in,height=1.25in,clip,keepaspectratio]{WBH}}]{Wael Boukley Hasan}
	received his MSc in Mobile Communications Engineering with distinction from Heriot-Watt University in 2013. After graduating, he worked for one year in Alcatel-Lucent in the Small Cells Platform Development Department. He then joined the CDT in Communications at the University of Bristol in 2014. His research within the CSN group is focused on investigating and developing different techniques for massive \gls{MIMO}, with an interest in increasing spectral efficiency and power efficiency. He was a member of the University of Bristol research team that set spectral efficiency world records in 2016 in collaboration with the research team from Lund University.
\end{IEEEbiography}
\begin{IEEEbiography}[{\includegraphics[width=1in,height=1.25in,clip,keepaspectratio]{Bio_Pics/Liang.jpg}}]{Liang Liu}
	received his B.S. and Ph.D. degree in the Department of Electronics Engineering (2005) and Micro-electronics (2010) from Fudan University, China. In 2010, he was with Rensselaer Polytechnic Institute (RPI), USA as a visiting researcher. He joined Lund University as a Post-doc in 2010. Since 2016, he is Associate Professor at Lund University. His research interest includes wireless communication system and digital integrated circuits design. He is a board member of the IEEE Swedish SSC/CAS chapter. He is also a member of the technical committees of VLSI systems and applications and CAS for communications of the IEEE circuit and systems society.
\end{IEEEbiography}
\begin{IEEEbiography}
	[{\includegraphics[width=1in,height=1.25in,clip,keepaspectratio]{mabeach}}]{Mark Beach}
received his PhD for research addressing the application of Smart Antenna techniques to GPS from the University of Bristol in 1989, where he subsequently joined as a member of academic staff. He was promoted to Senior Lecturer in 1996, Reader in 1998 and Professor in 2003. He was Head of the Department of Electrical \& Electronic Engineering from 2006 to 2010, and then spearheaded Bristol's hosting of the EPSRC Centre for Doctoral Training (CDT) in Communications. He currently manages the delivery of the CDT in Communications, leads research in the field of enabling technologies for the delivery of 5G and beyond wireless connectivity, as well as his role as the School Research Impact Director. Mark's current research activities are delivered through the Communication Systems and Networks Group, forming a key component within Bristol's Smart Internet Lab. He has over 25 years of physical layer wireless research embracing the application of Spread Spectrum technology for cellular systems, adaptive or smart antenna for capacity and range extension in wireless networks, MIMO aided connectivity for through-put enhancement, Millimetre Wave technologies as well as flexible RF technologies for SDR modems underpins his current research portfolio.
\end{IEEEbiography}
\begin{IEEEbiography}[{\includegraphics[width=1in,height=1.25in,clip,keepaspectratio]{SMD}}]{Simon Armour}
Simon Armour received the B.Eng. degree in Electronics and Communication Engineering from the University of Bath, Bath, U.K., in 1996, and the Ph.D. degree in Electrical and Electronic Engineering from the University of Bristol, Bristol, U.K., in 2001. Since 2001, he has been a Member of Academic Staff with the University of Bristol, where, since 2007, he has been a Senior Lecturer. He has authored or co-authored over 100 papers in the field of baseband Pysical layer and Medium Access Control layer techniques for wireless communications with a particular focus on \gls{OFDM}, coding, \gls{MIMO}, and cross-layer multiuser radio resource management strategies. He has investigated such techniques in general terms, as well as specific applications to Wireless Local Area Networks and cellular networks.
\end{IEEEbiography}
\begin{IEEEbiography}[{\includegraphics[width=1in,height=1.25in,clip,keepaspectratio]{Bio_Pics/OveE.jpg}}]{Ove Edfors}
	is Professor of Radio Systems at the Department of Electrical and Information Technology, Lund University, Sweden.
	His research interests include statistical signal processing and low-complexity algorithms with applications in wireless communications. 
	In the context of Massive MIMO, his main research focus is on how realistic propagation characteristics 
	influence system performance and base-band processing complexity.
\end{IEEEbiography}






\end{document}

%% file: acronyms.tex
\newacronym{4G}{4G}{fourth generation}
\newacronym{ADC}{ADC}{analogue-to-digital converter}
\newacronym{ADS}{ADS}{advanced design system}
\newacronym{AoA}{AoA}{angle of arrival}
\newacronym{AoD}{AoD}{angle of departure}
\newacronym{AP}{AP}{Access Point}
\newacronym{AS}{AS}{angular spread}
\newacronym{AWGN}{AWGN}{Additive White Gaussian Noise}
\newacronym{BBU}{BBU}{Base Band Unit}
\newacronym{BCC}{BCC}{Bristol City Council}
\newacronym{BER}{BER}{bit error rate}
\newacronym{BF}{BF}{beamforming}
\newacronym{BIO}{BIO}{Bristol Is Open}
\newacronym{BS}{BS}{base Station}
\newacronym{BW}{BW}{bandwidth}
\newacronym{CCDF}{CCDF}{complementary cumulative distribution function}
\newacronym{CDF}{CDF}{cumulative distribution function}
\newacronym{CDMA}{CDMA}{Code Division Multiple Access}
\newacronym{CDT}{CDT}{Centre for Doctoral Training}
\newacronym{COTS}{COTS}{Commercial off-the-shelf}
\newacronym{CP}{CP}{Cyclic Prefix}
\newacronym{CPA}{CPA}{coordinated pilot assignment}
\newacronym{CRC}{CRC}{Cyclic Redundancy Check}
\newacronym{CSI}{CSI}{channel state information}
\newacronym{CSIRO}{CSIRO}{Commonwealth Scientific and Industrial Research Organisation}
\newacronym{CSN}{CSN}{Communication Systems \& Networks}
\newacronym{D2D}{D2D}{device-to-device}
\newacronym{DAC}{DAC}{Digital-to-Analogue Converter}
\newacronym{dB}{dB}{decibels}
\newacronym{dBm}{dBm}{decibel-milliwatts}
\newacronym{DL}{DL}{downlink}
\newacronym{DMA}{DMA}{direct memory access}
\newacronym{DRAM}{DRAM}{dynamic random access memory}
\newacronym{DSP}{DSP}{Digital Signal Processing}
\newacronym{EM}{EM}{Electro-Magnetic}
\newacronym{EPSRC}{EPSRC}{Engineering and Physical Sciences Research Council}
\newacronym{ESD}{ESD}{Electro-Static Discharge}
\newacronym{FFT}{FFT}{Fast Fourier Transform}
\newacronym{FDD}{FDD}{frequency division duplex}
\newacronym{FD-MIMO}{FD-MIMO}{Full Dimension \gls{MIMO}}
\newacronym{FIFO}{FIFO}{first-in first-out}
\newacronym{FITRA}{FITRA}{Fast Iterative Truncation Algorithm}
\newacronym{FOC}{FOC}{Frequency Offset Correction}
\newacronym{FPGA}{FPGA}{field-programmable gate array}
\newacronym{FXP}{FXP}{fixed-point}
\newacronym{GPS}{GPS}{global positioning system}
\newacronym{GPSDO}{GPSDO}{\gls{GPS} disciplined oscillator}
\newacronym{GNSS}{GNSS}{Global Navigation Satellite Systems}
\newacronym{H2T}{H2T}{Host-to-target}
\newacronym{HD}{HD}{High-Definition}
\newacronym{HDL}{HDL}{Hardware Description Language}
\newacronym{ICT}{ICT}{Information and Communications Technology}
\newacronym{ID}{ID}{identity}
\newacronym{IFFT}{IFFT}{Inverse Fast Fourier Transform}
\newacronym{iid}{iid}{independent and indentically distributed}
\newacronym{IO}{IO}{Input/Output}
\newacronym{IP}{IP}{Intellectual Property}
\newacronym{ISI}{ISI}{Inter-Symbol Interference}
\newacronym{IUI}{IUI}{inter-user interference}
\newacronym{JSDM}{JSDM}{Joint Spatial Division Multiplexing}
\newacronym{KPH}{KPH}{kilometres per hour}
\newacronym{LDPC}{LDPC}{low-density parity-check}
\newacronym{LED}{LED}{Light Emitting Diode}
\newacronym{LF}{LF}{Loading Factor}
\newacronym{LIDAR}{LIDAR}{Laser Illuminated Detection \& Ranging}
\newacronym{LO}{LO}{local oscillator}
\newacronym{LOS}{LOS}{line-of-sight}
\newacronym{LS}{LS}{least squares}
\newacronym{LSD}{LSD}{Local Search Detection}
\newacronym{LTE}{LTE}{long-term evolution}
\newacronym{LUT}{LUT}{Look-up Table}
\newacronym{MAC}{MAC}{Medium Access Control}
\newacronym{MF}{MF}{matched filtering}
\newacronym{MRT}{MRT}{maximal ratio transmission}
\newacronym{MCS}{MCS}{modulation and coding scheme}
\newacronym{MGS}{MGS}{modified gram schmidt}
\newacronym{MIMO}{MIMO}{multiple-input, multiple-output}
\newacronym{ML}{ML}{Maximum likelihood}
\newacronym{MSE}{MSE}{Mean Square Error}
\newacronym{MMSE}{MMSE}{Minimum Mean Square Error}
\newacronym{MPD}{MPD}{Message Passing Detection}
\newacronym{MRC}{MRC}{maximal ratio combining}
\newacronym{MU}{MU}{multi-user}
\newacronym{MUI}{MUI}{Multi-User Interference}
\newacronym{MUSIC}{MUSIC}{Multiple Signal Classification}
\newacronym{MXI}{MXI}{Multisystem Extension Interface}
\newacronym{NI}{NI}{National Instruments}
\newacronym{NLOS}{NLOS}{non-line-of-sight}
\newacronym{OBR}{OBR}{Out-of-band power ratio}
\newacronym{OCXO}{OCXO}{Oven-controlled Crystal Oscillator}
\newacronym{OFDM}{OFDM}{Orthogonal Frequency Division Multiplexing}
\newacronym{OTA}{OTA}{over-the-air}
\newacronym{PA}{PA}{Power Amplifier}
\newacronym{PAPR}{PAPR}{peak-to-average power ratio}
\newacronym{PAS}{PAS}{Power-Azimuth Spectrum}
\newacronym{PC}{PC}{Power Control}
\newacronym{PCIe}{PCIe}{peripheral component interconnect express}
\newacronym{PDP}{PDP}{Power Delay Profile}
\newacronym{PDSCH}{PDSCH}{Physical Downlink Shared Channel}
\newacronym{PHY}{PHY}{physical layer}
\newacronym{PL}{PL}{Path-Loss}
\newacronym{PLL}{PLL}{Phase-locked Loop}
\newacronym{PMP}{PMP}{precoding, modulation and \gls{PAPR} reduction}
\newacronym{POC}{POC}{proof-of-concept}
\newacronym{P2P}{P2P}{Peer-to-peer}
\newacronym{ppb}{ppb}{parts-per-billion}
\newacronym{PPC}{PPC}{Pilot Power Control}
\newacronym{PSS}{PSS}{primary synchronisation signal}
\newacronym{PTS}{PTS}{Pilot Time Slot}
\newacronym{PUSCH}{PUSCH}{Physical Uplink Shared Channel}
\newacronym{PXIe}{PXIe}{\gls{PCIe} eXtensions for instrumentation}
\newacronym{QAM}{QAM}{quadrature amplitude modulation}
\newacronym{QPSK}{QPSK}{quadrature phase-shift keying}
\newacronym{QRD}{QRD}{QR decomposition}
\newacronym{RAN}{RAN}{Radio Access Network}
\newacronym{RB}{RB}{resource block}
\newacronym{R&D}{R\&D}{Research and development}
\newacronym{RE}{RE}{Resource Element}
\newacronym{RF}{RF}{radio frequency}
\newacronym{RIO}{RIO}{reconfigurable input/output}
\newacronym{RRH}{RRH}{remote radio head}
\newacronym{RSS}{RSS}{Received Signal Strength}
\newacronym{RT}{RT}{Ray-tracing}
\newacronym{RTM}{RTM}{Rear Transition Module}
\newacronym{RX}{RX}{Receive}
\newacronym{SC}{SC}{Single-Carrier}
\newacronym{SDMA}{SDMA}{Space-Divison Multiple Access}
\newacronym{SDR}{SDR}{software-defined radio}
\newacronym{SER}{SER}{Symbol Error Rate}
\newacronym{SINR}{SINR}{signal to interference plus noise ratio}
\newacronym{SIR}{SIR}{Signal to Interference Ratio}
\newacronym{SM}{SM}{Spatial Modulation}
\newacronym{SMA}{SMA}{SubMiniature version A}
\newacronym{SNR}{SNR}{signal to noise ratio}
\newacronym{SPDT}{SPDT}{Single Pole Double Throw}
\newacronym{SISO}{SISO}{Single-Input Single-Output}
\newacronym{SVD}{SVD}{singular value decomposition}
\newacronym{SVS}{SVS}{singular value spread}
\newacronym{T2H}{T2H}{Target-to-host}
\newacronym{TCF}{TCF}{time correlation function}
\newacronym{TDD}{TDD}{time division duplex}
\newacronym{TDOA}{TDOA}{Time-Difference-of-Arrival}
\newacronym{TO}{TO}{Timing Offset}
\newacronym{TPC}{TPC}{Transmission Power Control}
\newacronym{TR}{TR}{tone-reservation}
\newacronym{TX}{TX}{Transmit}
\newacronym{TTI}{TTI}{Transmission Time Interval}
\newacronym{U8}{U8}{Unsigned 8-bit}
\newacronym{UDN}{UDN}{ultra-dense network}
\newacronym{UDP}{UDP}{User Datagram Protocol}
\newacronym{UE}{UE}{user equipment}
\newacronym{UI}{UI}{User Interface}
\newacronym{UL}{UL}{uplink}
\newacronym{USRP}{USRP}{universal software radio peripheral}
\newacronym{UoB}{UoB}{University of Bristol}
\newacronym{VLC}{VLC}{VideoLAN Client}
\newacronym{VTC}{VTC}{Vehicular Technology Conference}
\newacronym{WARP}{WARP}{Wireless Open-Access Research Platform}
\newacronym{ZF}{ZF}{zero-forcing}